\documentclass[letterpaper,pra,twocolumn,showpacs]{revtex4}
\usepackage{graphicx,bm,color}    

\begin{document}
\title{Possible quantum phase-manipulation of a two-leg ladder in mixed-dimensional fermionic cold atoms}
\author{Wen-Min Huang, Kyle Irwin and Shan-Wen Tsai}
\affiliation{Department of Physics and Astronomy, University of California, Riverside, CA 92521, USA.}

\date{\today}


\begin{abstract}
The recent realization of mixed-dimensional systems of cold atoms has attracted much attention from both experimentalists and theorists. Different effective interactions and novel correlated quantum many-body phases may be engineered in these systems, with the different phases being tunable via external parameters. In this article we investigate a two-species Fermi atom mixture: one species of atom exists in two hyperfine states and is confined to move in a two-leg ladder, interacting with an on-site interaction, and the other moves freely in a two dimensional square lattice that contains the two-leg ladder. The two species of atoms interact via an on-site interaction on the ladder. In the limit of weak inter-species interactions, the two-dimensional gas can be integrated out, leading to an effective long-range mediated interaction in the ladder, generated by to the on-site inter-species interaction. We show that the form of the mediated interaction can be controlled by the density of the two-dimensional gas and that it enhances the  charge density wave instability in the two-leg ladder after the renormalization group transformation. Parameterizing the phase diagram with various experimentally controllable  quantities, we discuss the possible tuning of the macroscopic quantum many-body phases of the two-leg ladder in this mixed-dimensional fermionic cold atom system. 
\end{abstract}

\maketitle

Over the past years, Fermi mixtures in ultracold atoms have been achieved experimentally\cite{ZwierleinS,PartridgeS,ZwierleinN,ShinP}, and have triggered intense studies by experimentalists and theorists alike. A  multitude of exotic phenomena, which normally do not occur in condensed matter systems, can now be fabricated with exceptional tunable parameters\cite{Jakscha,Bloch05,Taglieber,Wille,Voigt,Spiegelhalder,Tiecke10}. For instance, fermion mixtures can be manipulated to carry different internal properties such as population densities, group symmetries, lattice geometries, and dimensionalities. By cooling two fermionic isotopes of ytterbium with different nuclear spins, a degenerate Fermi mixture with an SU(2)$\times$SU(6) symmetry has recently been experimentally achieved\cite{Taie10}. Mixed two-species Fermi gases with unequal populations have also been extensively studied\cite{PartridgeP,ShinN}, as well as systems of dipolar atoms and molecules\cite{dipolar1,dipolar2,dipolar3}.
Many-body effects in these systems
can be further enhanced when the system is loaded onto an optical lattice\cite{Bloch08} or via Feshbach resonance\cite{Chin,Timmermans,Trenkwalder}. Fermions with imbalanced populations\cite{Lai} and dipolar fermions\cite{Bhongale1,Bhongale2} on a square lattice have been investigated and present a much richer phase diagram than their gas counterparts, with lattice effects enhancing the transition temperature for various phases.

Experimentally, it is also possible to confine components of a ultracold atom mixture in different dimensions\cite{Lamporesi10,Haller10}, and this has trigged some recent theoretical studies\cite{Nishida08,Nishida101,Iskin}. 
As pointed out by Nishida\cite{Nishida10}, the mediated intra-species interaction generated by the interaction between species moving in different dimensions may reshape the phase diagram in a bilayer Fermi gas. Furthermore, Efimov physics with fermions is also proposed to emerge in one/three-dimensional mixed systems\cite{Nishida09}. The properties of the lower-dimensional  Fermi gas can be manipulated by tuning parameters in the higher dimensional species. For example, the phase diagram of a single chain is strongly modified by being embedded in a two-dimensional lattice\cite{Irwin}.

\begin{figure}[t]
\begin{center}
\includegraphics[width=7cm]{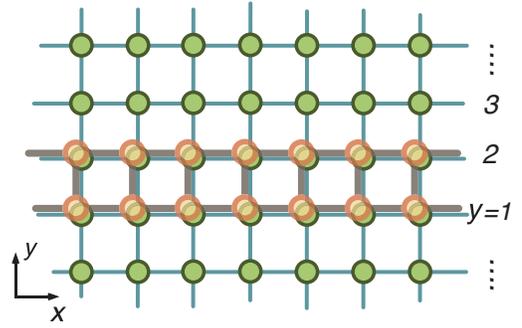}
\caption{(Color online)A two-leg ladder embedded in a 2D square lattice and the coordinate system are illustrated.  }
\label{lattice}
\end{center}
\end{figure}

Inspired by these experiments, we study a mixed-dimensional two-species fermionic system: one species confined in a two-leg ladder with on-site repulsion, the other moving freely in a two-dimensional (2D) square lattice. 
An inter-species interaction is also introduced as on-site due to the energy cost of double occupation. 
Integrating out the 2D fermionic gas, a mediated long-range interaction is generated in the ladder. We show that the mediated interaction moves the bare on-site interaction in the two-leg ladder off the symmetric point, and enhances the charge density wave (CDW) instability in the renormalization group (RG) transformation. Here we use the term charge-density-wave to refer to density modulations in analogy with nomenclature used in the study of electronic systems, even though the atoms are charge neutral. We find that, by controlling the filling in the 2D gas, the phase of the ladder can be tuned. By mapping out the phase diagram for various parameters, we show the possible quantum phase-manipulation of a two-leg ladder in mixed dimensional fermionic cold atoms.

A schematic of the system we consider here is illustrated in Fig.~\ref{lattice}. 
The action can be written as $S=S_l+S_c^0+S_{cl}$, where $S_l$ stands for the action for the two-leg ladder with on-site interaction $U_l$. The action of the non-interacting 2D system is denoted as
\begin{eqnarray}
S^0_{c}=\int d\tau\sum_{\langle\bm{r},\bm{r}'\rangle}\bar{\phi}_{\alpha}(\bm{r},\tau)\left[\partial_{\tau}\delta_{\bm{r},\bm{r}'}+H_{c}(\bm{r},\bm{r}')\right]\phi_{\alpha}(\bm{r}',\tau).
\end{eqnarray}
where $\bar{\phi}_{\alpha}$,$\phi_{\alpha}$ are Grassmann fields with (pseudo)spin index $\alpha$, and the Hamiltonian is represented as $H_{c}(\bm{r},\bm{r}')=\left[t_c(\bm{r},\bm{r}')-\mu_{2D}\delta_{\bm{r},\bm{r}'}\right]$ with chemical potential $\mu_{2D}$. 
Moreover, the uniform hopping amplitude $t_c(\bm{r},\bm{r}')=-1$ when $\bm{r}$ and $\bm{r}'$ represent nearest neighbor sites. The energy cost of for overlap of two atoms of different species can be regarded as the on-site inter-species repulsion with strength $U_{cl}$ and represented as
\begin{eqnarray}
S_{cl}=\int d\tau\hspace{0.1cm} U_{cl}\sum_{a}\sum_{\bm{r}}n_c(\bm{r},\tau)n_l(\bm{r},\tau)\delta_{y,a},
\end{eqnarray}
\noindent where $n_c=\sum_{\alpha}\bar{\phi}_{\alpha}(\bm{r})\phi_{\alpha}(\bm{r})$ and $n_l=\sum_{\alpha}\bar{\varphi}_{y\alpha}(x)\varphi_{y\alpha}(x)$ are the densities for the 2D lattice and the ladder, respectively. The summation over $a=1,2,\cdots,N$ stands for the position of $N$-legs along the $y$-direction, as shown as Fig.~\ref{lattice}. Here we focus on the $N=2$ case.

Considering the limit of weak inter-species interactions, we expand the action in powers of  $U_{cl}$ and integrate out the non-interacting 2D gas, neglecting terms $O(U_{cl}^3)$. The first 
term in the expansion gives a correction to the chemical potential of the ladder system. Since this only slightly shifts the phase boundaries, we ignore it here. The second term generates a mediated interaction, modifying the profile of the total effective interaction in the two-leg ladder. The action for the ladder can now be written as $S_{\rm eff} = S_l + S_{\rm med}$, where
\begin{eqnarray}\label{Seff}
\nonumber S_{\rm med}&&\hspace{-0.3cm}\simeq\int d\tau_1\int d\tau_2\hspace{0.1cm} \frac{U^2_{cl}}{2}\hspace{0.1cm}\sum_{a,b}\delta_{y_1,a}\delta_{y_2,b}\hspace{0.1cm}{\rm tr}\Big[n_l(\bm{r}_1,\tau_1)\\
&&\hspace{-0.5cm}\mathcal{G}_0(\bm{r}_1,\tau_1;\bm{r}_2,\tau_2)n_l(\bm{r}_2,\tau_2)\mathcal{G}_0(\bm{r}_2,\tau_2;\bm{r}_1,\tau_1)\Big],
\end{eqnarray}
with $\mathcal{G}_0(\bm{r},\tau;\bm{r}',\tau')=\left[\partial_{\tau}\delta_{\bm{r},\bm{r}'}-H_{c}(\bm{r},\bm{r}')\right]^{-1}$ the imaginary-time Green's function. This shows that a mediated interaction between particle $n_{l}(\bm{r}_1,\tau_1)$ and $n_{l}(\bm{r}_2,\tau_2)$ with retardation effects is generated in the ladder system. Retardation effects can be neglected when the Fermi velocity of the 2D fermions is large compared with the one for the ladder. 

We now study the effects of the effective interaction on the ladder system and determine its phase diagram using a RG technique. 
By ignoring retardation effects, the Grassmann number $\varphi$ is  decomposed into chiral pairs\cite{Balents96,Lin97,Lin98,Lin05,Bundler08,Bundler09}, $\varphi_{y}(x)\approx\sum_{P,i}T_{yi}\psi_{Pn}(x)e^{iPk_{F_i}x}$, where $P=R/L=+/-$ represent right/left-moving particles, and $k_{F_i}$ is the Fermi wavelength of the band index $i$. The transformation matrix between leg-index $y$ and band-index $i$ in the $N$-ladder is introduced as $T_{yi}=\sqrt{\frac{2}{N+1}}\sin\left[\frac{\pi}{N+1}yi\right]$\cite{Lin98}. The interactions between these chiral fermions can be categorized as Cooper scatterings $c^l_{ij}$, $c^s_{ij}$ and forward scatterings $f^l_{ij}$, $f^s_{ij}$, where we set $f_{ii}=0$ since $f_{ii}=c_{ii}$\cite{Lin05}. Thus, the mediated interactions in a ladder system can be written as
\begin{eqnarray}
\nonumber \hspace{-0.4cm}S_{\rm med}\hspace{-0.3cm}&&\simeq\frac{U_{cl}^2}{U_l}\int d\tau \int dx  \\
&&\nonumber \hspace{-0.4cm}\Big[f^s_{ij}\bar{\psi}_{Ri\alpha}\psi_{Ri\alpha}\bar{\psi}_{Lj\beta}\psi_{Lj\beta}+f^l_{ij}\bar{\psi}_{Ri\alpha}\psi_{Lj\alpha}\bar{\psi}_{Lj\beta}\psi_{Ri\beta}\\
&&\hspace{-0.6cm}+c^s_{ij}\bar{\psi}_{Ri\alpha}\psi_{Rj\alpha}\bar{\psi}_{Li\beta}\psi_{Lj\beta}+c^l_{ij}\bar{\psi}_{Ri\alpha}\psi_{Lj\alpha}\bar{\psi}_{Li\beta}\psi_{Rj\beta}\Big],
\end{eqnarray}
where the bare values of the couplings are expressed as
\begin{eqnarray}\label{BareC}
\nonumber&&\hspace{-0.8cm}f^s_{ij}=U^{\rm med}_{iijj}(0), \ \ f^l_{ij}=U^{\rm med}_{ijji}(k_{F_i}+k_{F_j}),\\
&&\hspace{-0.8cm}c^s_{ij}=U^{\rm med}_{ijij}(k_{F_i}-k_{F_j}), \ \ c^l_{ij}=U^{\rm med}_{ijij}(k_{F_i}+k_{F_j}),
\end{eqnarray}
with the definition of dimensionless mediated interactions, $U^{\rm med}_{ijkl}(k)=U_l\int^{\pi}_{-\pi}\frac{dq}{2\pi}\hspace{0.1cm}\Gamma_{ijkl}(q)\chi_0(k,q)$.
Here the particle-hole propagator is defined as
\begin{eqnarray}
\chi_0(k,q)\hspace{-0.05cm}=\hspace{-0.15cm}\int\hspace{-0.1cm}\frac{d\bm{p}}{4\pi^2}\frac{n_{F}\Big[\epsilon(\bm{p})\Big]\hspace{-0.1cm}-\hspace{-0.05cm}n_{F}\Big[\epsilon(p_x+k,p_y+q)\Big]}{\epsilon(\bm{p})-\epsilon(p_x+k,p_y+q)+i0^{+}},
\end{eqnarray}
coming from the convolution of the two momentum-space Green's functions in Eq.~(\ref{Seff}), where $\epsilon(\bm{p})=-2(\cos p_x+\cos p_y)-\mu_{2D}$ is the dispersion of the 2D gas and $n_{F}$ is the Fermi-Dirac distribution. Furthermore, the extra kernel in the mediated interactions is defined as $\Gamma_{ijkl}(q)=\sum_{a,b}e^{iq(b-a)}T^*_{ai}T_{aj}T^*_{bk}T_{bl}$, summing over the leg indices $a,b=1,2,\cdots,N$. It is worthwhile to notice that the mediated interactions are the Ruderman-Kittel-Kasuya-Yoshida (RKKY) type\cite{Ashcroft}. However, the exact profile in real space can not be computed analytically since we consider a lattice model.  

In the absence of the mediated interaction, all bare values of the couplings in RG equations would be the same since only on-site interactions are considered. In the presence of the mediated interactions, this symmetry is broken and the initial couplings renormalize in very different ways, as shown as Eq.~(\ref{BareC}). However, since the mediated interactions are determined only by the exchanged momenta in spin-conserving scattering processes, couplings sharing the same exchanged momenta remain the same bare values. For instance, zero momenta is exchanged in both $c_{ii}^{s}$ and $f_{ij}^s$ and they therefore have the same bare value, similarly, for $c_{ij}^l$ and $f_{ij}^l$ with exchanged momenta $(k_{F_i}+k_{F_j})$.



The RG equations of a $N$-leg ladder can be found in the literature\cite{Seidel}, and are only solved numerically. After taking the effective interactions as the initial condition and integrating the differential RG equations, the flows of the couplings can be obtained. By analyzing these flows with the scaling Ansatz, $g_{i}\sim 1/\left(l_d-l\right)^{\gamma_{i}}$, where $l_d$ is the divergent length scale in one-loop RG, the hierarchy of the relevant couplings can be directly read out from the RG exponent $\gamma_i$\cite{Shih,Cai}. Combining with the Abelian bosonization method, the phase diagram of a ladder system is determined\cite{Lin98,Bundler08,Bundler09}. 
Furthermore, the relative values of the charge and spin gaps between different Fermi points can be determined by the RG exponents\cite{Shih}. This allows us to distinguish the $d$-wave superconductivity with an anisotopic spin-gap (referred as $d$-SC$_2$ here) in a two-leg ladder with heavy doping, $n<0.6$. 
We note that there is not a real phase transition between the $d$-wave superconductivity with isotropic spin gaps ($d$-SC) and $d$-SC$_2$, because the number of spin and charge gaps and relative sign of relevant couplings are the same. However, it is useful to emphasize this difference here because the anisotropic spin gaps can measured, and therefore the two phases can be distinguished experimentally.

\begin{figure}[t!]
\begin{center}
\includegraphics[width=8.cm]{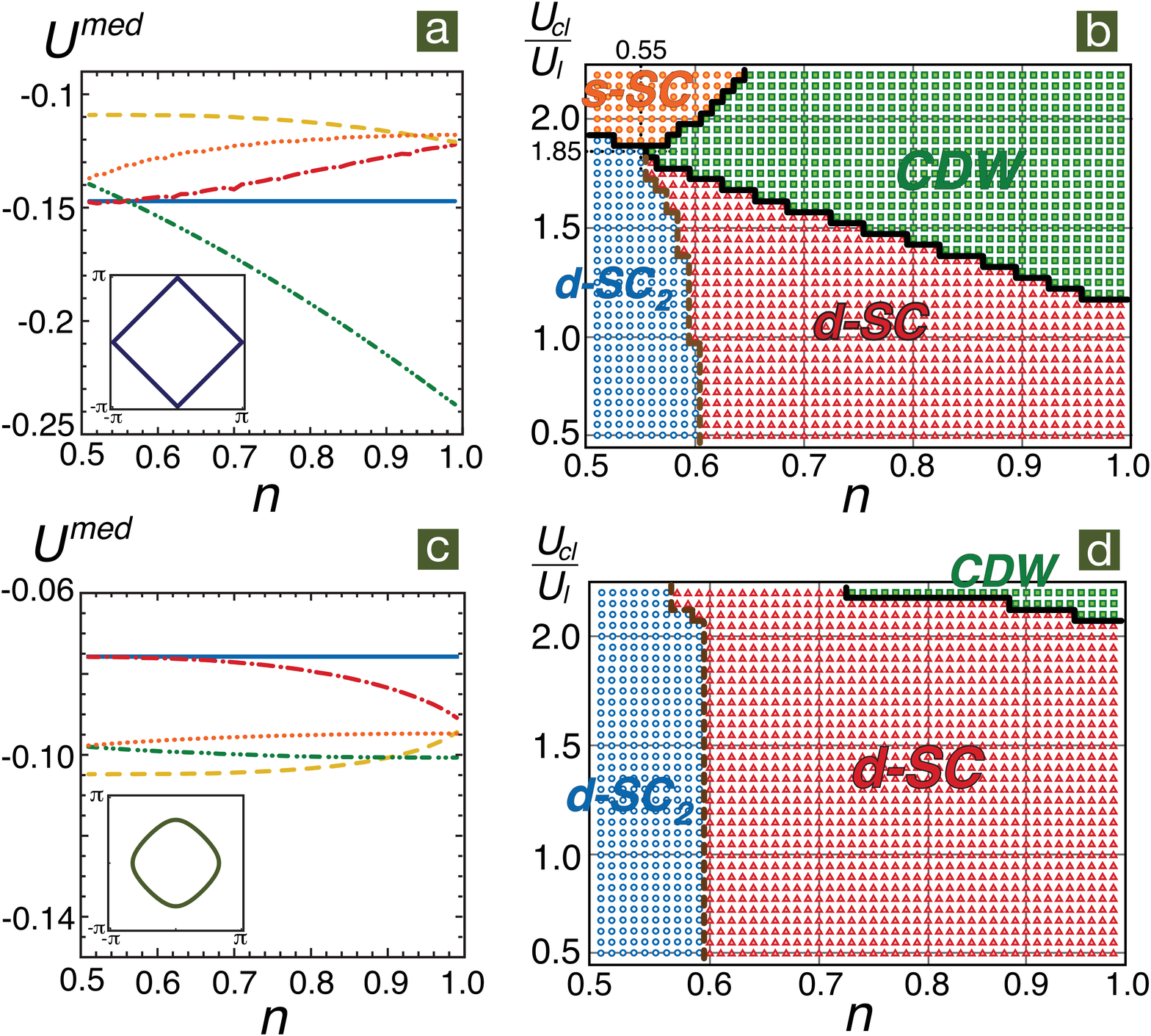}
\caption{(Color online) Interactions mediated by a 2D gas with (a) $\mu_{2D}=0$ and (c) $\mu_{2D}=1$ versus density $n$ of the ladder: The blue solid line represents $c^{s}_{11}$, $c^{s}_{22}$ and $f^{s}_{12}$. $c_{12}^l$ and $f^{l}_{12}$ are denoted as the green dot-dash-dotted line. The red dot-dashed, orange dotted, and brown dashed lines denote $c_{11}^l$, $c_{12}^s$ and $c_{22}^l$ respectively. The corresponding phase diagrams for (a) and (c) are illustrated in (b) and (d), respectively.}
\label{PDmu}
\end{center}
\end{figure}

%
To illustrate how the mediated interaction depends on the filling of the 2D fermions, we show results for $\mu_{2D}=0$, when the 2D Fermi surface is nested (Fig.~\ref{PDmu}a), and for $\mu_{2D}=1$ (Fig.~\ref{PDmu}c). The corresponding phase diagrams are shown in Fig.~\ref{PDmu}b and Fig.~\ref{PDmu}d.
In Fig.~\ref{PDmu}b, when the strength of the coupling $U_{cl}/U_{l}$ is larger than $1.2$, a CDW starts to emerge from $d$-SC near half-filling ($n\simeq1$). Increasing $U_{cl}/U_{l}$ brings the phase boundary to higher hole-doping. It is interesting to note that the Luttinger liquid phase that appears near $n\gtrsim0.5$ for the standard Hubbard model\cite{Lin98} is destroyed by the mediated interactions. We find a three-phase ``triple-point'' (CDW, $d$-SC and $s$-SC) at $U_{cl}/U_{l}=1.85$ and $n=0.55$. From there, increasing the $U_{cl}/U_{l}$ ratio gives $c_{11}^{l}(0)$ (or $c_{11}^{\sigma}(0)$) negative, and an $s$-SC phase occurs, as expected by BCS theory. 
Doping away from the ``triple-point", $f_{12}^{\sigma}(0)$ becomes negative and CDW dominates. Roughly speaking, when $U_{cl}/U_{l}>1.85$, the phase will be determined by the competition between the negative values of $f_{12}^{\sigma}(0)$ and $c_{11}^{\sigma}(0)$.

It is important to point out that the CDW phase emerges due to enhanced mediated interactions $f^{l}_{12}$ near half-filling. When $n\simeq1$, the momentum exchanged during $f^{l}_{12}$ processes approaches $\pi$. The particle-hole propagator in the mediated interaction shows a maximum contribution at $\mu_{2D}=0$, resulting from  maximum electron-hole pairing [from $\cos p_x$ and $\cos(p_x+\pi)=-\cos p_x$] in the 2D dispersion along the $x$-direction. Moving away from $n=1$, the mediated interaction of $f_{12}^{l}$ decreases, at which point the CDW phase ceases to occur. 
When we increase the 2D chemical potential, the same behavior occurs. As shown in Fig.~\ref{PDmu} (d), the regime of the CDW phase shrinks at $\mu_{2D}=1$ resulting from weak mediated interactions. In this situation the mediated interactions vary smoothly, and show little effect in the ladder system. 
Therefore, the mediated interaction depends strongly on the chemical potential in the 2D system, and consequently, the phase diagram of a two-leg ladder can be modified/tuned via the filling in the 2D square lattice. 

\begin{figure}[t]
\begin{center}
\includegraphics[width=8.cm]{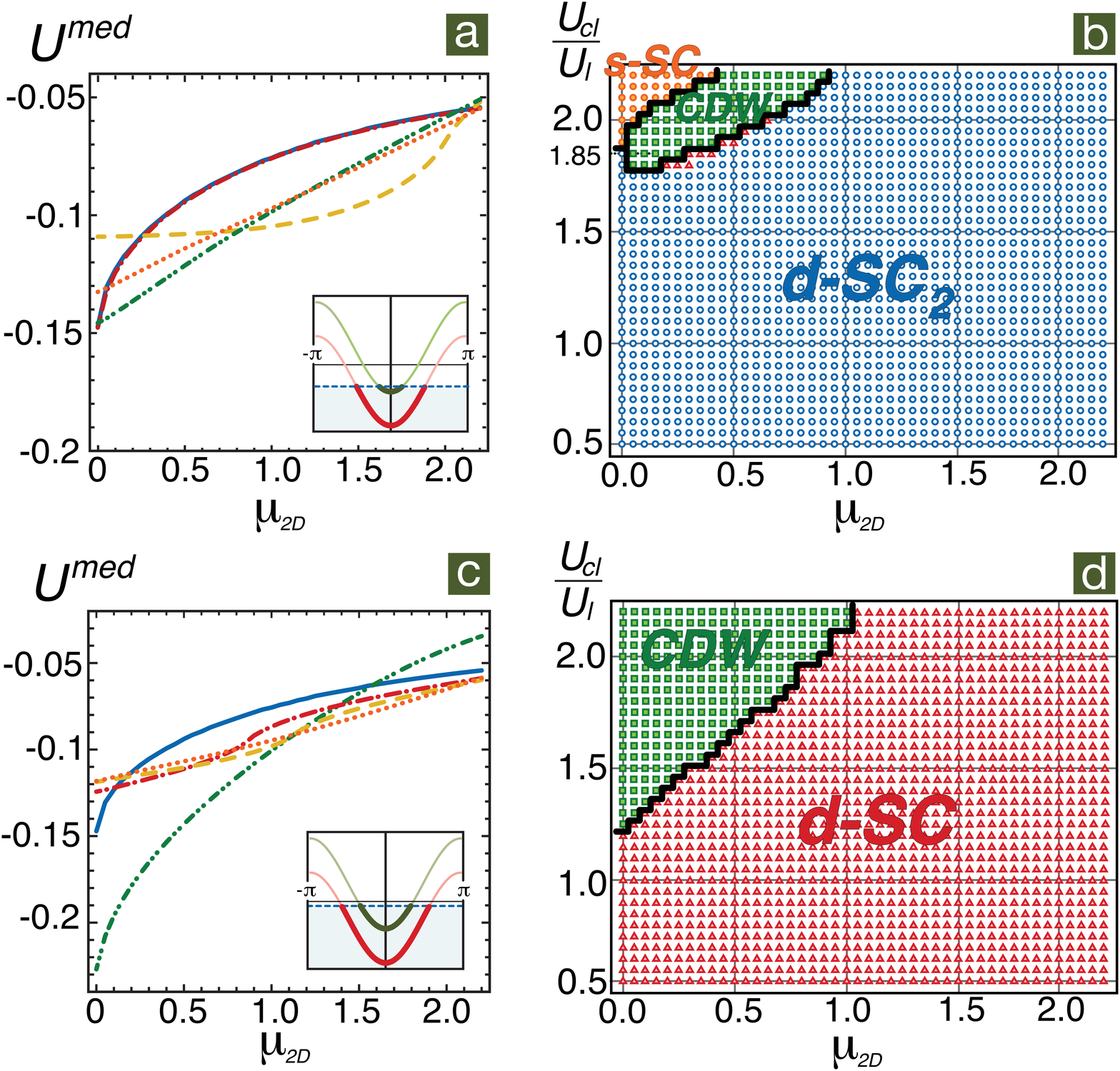}
\caption{(Color online)The mediated interactions from a 2D gas with densities $n=0.55$ and $n=0.95$ on the ladder are plotted as function of the chemical potential $\mu_{2D}$ of the 2D system (in units of the 2D hopping amplitude $t_c$) in (a) and (c), respectively. The strokes for couplings in (a) and (c) are the same as in Fig.~\ref{PDmu}. Corresponding phase diagrams are shown in (b) and (d).}
\label{PDn}
\end{center}
\end{figure}

To further illustrate the influence of the 2D density, we show the mediated interactions and corresponding phase diagrams parameterized by the chemical potential in the 2D system in Fig.~\ref{PDn} (a)(b) and (c)(d) for fixed density in the ladder, $n=0.55$ and $n=0.95$, respectively. When $U_{cl}/U_l$ is larger than $1.2$ at $n=0.95$, the phase transition between CDW and $d$-SC happens near $\mu_{2D}=0$. However, the phase boundary moves to higher values of $U_{cl}/U_l$ when the chemical potential $\mu_{2D}$ increases
and the mediated interaction decreases.
In general, the tendency of the mediated interaction is to decrease upon the increasing of the filling in the 2D system (or $\mu_{2D}$). This is due to the non-interacting assumption used in the 2D gas. Under this condition, the strength of the mediated interactions is roughly proportional to the density near the Fermi surface, which is maximum at $\mu_{2D}=0$.  
  
Although the pattern of the mediated interactions in real space is rather complicated, it can be 
approximated as an effective on-site attraction and a nearest-neighbor repulsion, since longer range terms are rapidly decaying. The phase diagram can then be understood from these two effective interactions. By tuning the chemical potential in 2D, CDW emerges when the effective nearest-neighbor repulsion exceeds the original on-site repulsive interaction. However, increasing the $U_{cl}/U_l$ ratio makes the magnitude of the effective on-site attraction larger than the original on-site repulsion. When the average on-site attraction is weak, CDW and $s$-SC phases compete. Nevertheless, the $s$-SC phase is dominant for large $U_{cl}/U_l$, and the inter-species interaction can be regarded as a glue to pair the fermions, similar to the electron-phonon interaction in a conventional superconductor. 

In conclusion, we studied a two-species Fermi mixture composed of a species that moves in a two-leg ladder with on-site interaction, and another, non-interacting, species moving on a 2D square lattice. By integrating the 2D Fermi gas in the limit of weak inter-species coupling, a long-range mediated interaction is generated, and the CDW instability in the ladder is enhanced. We show the phase diagram parameterized by the filling in both the two-leg ladder and the 2D square lattice.

\textit{Acknowledgments - }
KI and SWT gratefully acknowledge support from NSF under Grant DMR-0847801 and from the UC-Lab FRP under Award number 09-LR-05-118602. Meanwhile, WMH sincerely acknowledges support from NSC Taiwan under Grant 101-2917-I-564-074.

\end{document}